\documentclass[twocolumn,showpacs,preprintnumbers,amsmath,amssymb,APSl,prd,nofootinbib,superscriptaddress]{revtex4-1}

\usepackage{dcolumn}
\usepackage{bm}
\usepackage{ifpdf}
\usepackage{hyperref}
\usepackage{dcolumn}
\usepackage{bm}
\usepackage[spanish,english]{babel}
\usepackage{amsfonts}
\usepackage{amssymb}
\usepackage{graphicx}
\usepackage[latin1]{inputenc}

\newcommand \be{\begin{equation}}
\newcommand \en{\end{equation}}
\newcommand \bea{\begin{eqnarray}}
\newcommand \ena{\end{eqnarray}}

\begin{document}

\title{Black holes in five-dimensional Palatini $f(R)$ gravity and implications for the AdS/CFT correspondence}

\author{D. Bazeia} \email{bazeia@fisica.ufpb.br}
\affiliation{Departamento de F\'isica, Universidade Federal da
Para\'\i ba, 58051-900 Jo\~ao Pessoa, Para\'\i ba, Brazil}
\author{L. Losano} \email{losano@fisica.ufpb.br}
\affiliation{Departamento de F\'isica, Universidade Federal da
Para\'\i ba, 58051-900 Jo\~ao Pessoa, Para\'\i ba, Brazil}
\author{Gonzalo J. Olmo} \email{gonzalo.olmo@csic.es}
\affiliation{Departamento de F\'isica, Universidade Federal da
Para\'\i ba, 58051-900 Jo\~ao Pessoa, Para\'\i ba, Brazil}
\affiliation{Departamento de F\'{i}sica Te\'{o}rica and IFIC, Centro Mixto Universidad de
Valencia - CSIC. Universidad de Valencia, Burjassot-46100, Valencia, Spain}
\author{D. Rubiera-Garcia} \email{drubiera@fisica.ufpb.br}
\affiliation{Departamento de F\'isica, Universidade Federal da
Para\'\i ba, 58051-900 Jo\~ao Pessoa, Para\'\i ba, Brazil}


\date{\today}

\begin{abstract}
We show that theories having second-order field equations in the context of higher-dimensional modified gravity are not restricted to the family of Lovelock Lagrangians, but can  also be obtained if no a priori assumption on the relation between the metric and affine structures of space-time is made (Palatini approach). We illustrate this fact by considering the case of Palatini $f(R)$ gravities  in five dimensions. Our results provide an alternative avenue to explore new domains of the AdS/CFT correspondence without resorting to {\it ad hoc} quasi-topological constructions.
\end{abstract}

\maketitle

\section{Introduction}

It is usually stated that in five dimensions only the Gauss-Bonnet (GB) theory provides second-order field equations. This is due to the fact that the coefficients of the higher-order curvature invariants $R^2$, $R_{\mu\nu}R^{\mu\nu}$ and $R_{\alpha\beta\gamma\delta}R^{\alpha\beta\gamma\delta}$ summing up the GB Lagrangian correspond to specific choices that remove the undesired fourth-order terms. Having second-order field equations in a theory of gravity is essential in order to obtain exact solutions and get rid of troubles with ghost-like instabilities. In particular, paging through the literature, one finds a large number of black hole solutions in different GB scenarios \cite{GB-solutions}. More curvature invariant terms can be added to the GB action, leading to the general family of Lovelock Lagrangians \cite{GB}, for which black hole solutions have been found in some particular cases \cite{Lovelock-solutions}. In this family, the cosmological constant term and the Einstein-Hilbert Lagrangian of General Relativity (GR) represent the zeroth-order and first-order terms, respectively, GB theory the second-order, and so on. However, in five (or six) dimensions, the extra terms beyond the GB combination turn out to be a topological invariant and therefore do not contribute to the field equations.

The \emph{quasi-topological} approach \cite{quasi-topological} takes another route to this problem by adding a set of new curvature-cubed terms to the Einstein-Hilbert action in five dimensions in such a way that, though the field equations are of third-order, their linearized counterpart describing gravitons propagating in an Anti-de Sitter background are second-order (so the resulting theory could be ghost-free). Besides its unnaturalness, this approach is troubled by the great difficulty of obtaining exact solutions.

The interest in having more terms added to the action, as in the Lovelock and quasi-topological gravities, can be traced back to the fact that, via the Anti-de Sitter/Conformal Field Theory (AdS/CFT) correspondence, new couplings in the gravitational action may broaden the class of dual CFTs that one can study using holographic methods \cite{AdSCFT}. These new terms also arise in the quantization of fields in curved space-time \cite{quantumHC} and in approaches to quantum gravity \cite{stringHC}.

Here we note that the reason explaining why GB (and Lovelock) gravity provides second-order field equations is the fact that it actually belongs to the class of Palatini theories \cite{Borunda08}. These theories are constructed removing any a priori relation of the connection with respect to the metric. Indeed, as metric and connection carry a very different geometrical meaning, the former defining the measurements of lengths, areas, volumes, etc., and the latter being related to the existence of properties remaining invariants under affine transformations (such as parallelism), they are conceptually different and unrelated a priori (see \cite{Zanelli} for a pedagogical discussion). The connection is usually taken to be metric-compatible due to the fact that the Palatini approach for GR (and, more generally, for  the Lovelock family) gives an additional equation for the connection whose solution is precisely the Levi-Civita one. Over the debate of using the minimum number of geometrical entities (the metric approach) or the minimum number of a priori hypothesis (the Palatini approach), the former has been traditionally favored. Indeed, the coincidence of the metric and Palatini formulation of GR, which is due to the particular functional form of the Einstein-Hilbert action, has led most researchers in the field to implicitly assume that the connection must always be compatible with the metric $g_{\mu\nu}$. This is, however, not necessarily true for nonlinear extensions of GR.

In general, the mathematical and physical structures of metric-affine (or Palatini) theories largely differ from their metric counterparts \cite{Borunda08b}. Now, a nice feature of Palatini gravities is that, for a large family of functional forms in the action, the metric field equations remain second-order, as in GR. The connection turns out to be a constrained object that can be solved in terms of the metric and the matter fields. This opens the door to new scenarios where the physics of second-order modified gravity theories can be explored, without unnecessary constraints on the coefficients of the theory and/or ad hoc constructions. In particular, theories including higher-order curvature invariants do not need to be constrained to have the same form as the GB/Lovelock one, which provides new avenues for studying exact solutions of gravity theories in the context of higher-dimensions and broadens the possibilities of the AdS/CFT route. Note that due to the fact that the GB combination is a topological invariant in four dimensions, whose variation does not provide modified dynamics as compared to GR, the natural framework for the comparison between solutions of metric and Palatini approaches is that of five (and six) dimensions.

In a series of papers \cite{or12} we have developed tools to solve Palatini field equations in four dimensions coupled to several kinds of matter and worked out exact solutions. Since the variational principle provides two sets of equations, the strategy to solve them consists on identifying a new rank-two tensor $h_{\mu\nu}$ as the metric for which the independent connection $\Gamma_{\mu\nu}^{\lambda}$ becomes the Levi-Civita one. In terms of $h_{\mu\nu}$ the field equations are second-order and, since $h_{\mu\nu}$ is algebraically related (via the matter sources) to the metric $g_{\mu\nu}$ appearing in the definition of the action, the field equations for $g_{\mu\nu}$ are second-order as well. The non-trivial role played by the matter sources becomes a distinctive feature of Palatini theories, giving rise to new effects such as the existence of wormhole geometries \cite{or12}, which can be supported even by elementary electric fields. In this work we shall consider the simplest scenario of Palatini gravity in extra dimensions, namely, five-dimensional $f(R)$ theories (see \cite{Olmo:2011uz} and \cite{fRr} for recent reviews) and explicitly show that the field equations remain second-order. Moreover, analytical solutions are obtained using electrovacuum fields. Discussion on the extension of this approach to the case of more general actions for the gravitational field is also provided.

\section{Theory and second-order field equations}

We consider the action of $f(R)$ gravity, defined as follows

\be \label{eq:action}
S=\frac{1}{2\kappa^2} \int d^n x \sqrt{-g} f(R) +S_m(g_{\mu\nu},\psi_m),
\en
where $n$ is the number of space-time dimensions, $\kappa^2$ is the $n$-dimensional Newton's constant in some appropriate system of units, $g$ is the determinant of the space-time metric $g_{\mu\nu}$, $f(R)$ is a given function of the curvature scalar $R=g^{\mu\nu}R_{\mu\nu}(\Gamma)$, where $\Gamma \equiv \Gamma_{\mu\nu}^{\lambda}$ is the independent connection, $S_m$ is the matter action and $\psi_m$ denotes collectively the matter fields. As we are working in the Palatini formalism, metric and connection are independent entities and, therefore, the variational principle must be applied to both of them. For simplicity, we assume vanishing torsion $\Gamma_{[\mu\nu]}^{\lambda}=0$ (see \cite{OR_torsion} for a discussion of the role of torsion in these theories). Under these conditions, the variation of the action (\ref{eq:action}) gives
\bea \label{eq:var-total}
\delta S&=& \frac{1}{2\kappa^2} \int d^n x  \sqrt{-g} \Big[ \Big(f_R R_{\mu\nu}-\frac{1}{2} f g_{\mu\nu}\Big) \delta g^{\mu\nu} \nonumber \\
&+& f_R g^{\mu\nu} \delta R_{\mu\nu}(\Gamma)\Big] + \delta S_m,
\ena
where we have used the short-hand notation $f_R \equiv df/dR$. Let us take care first of the connection piece. It can be rewritten, using $\delta R_{\mu\nu}= \nabla_{\lambda} \delta \Gamma_{\nu\mu}^{\lambda} - \nabla_{\nu} \delta \Gamma_{\lambda \mu}^{\lambda}$, as
\be \label{eq:var-con}
\delta_{\Gamma} S = \frac{1}{2\kappa^2} \int d^n x\sqrt{-g} f_R g^{\mu\nu} \Big(\nabla_{\lambda} \delta \Gamma_{\nu \mu}^{\lambda} - \nabla_{\nu} \Gamma_{\lambda \mu}^{\lambda} \Big).
\en
Integrating by parts we express (\ref{eq:var-con}) as
\bea\label{eq:con-var}
\delta_{\Gamma} S&=& \frac{1}{2\kappa^2} \int d^n x \sqrt{-g} \Big[- \nabla_{\lambda} \Big(\sqrt{-g} f_R g^{\mu\nu} \Big)  \\
&+& \frac{1}{2} \delta_{\lambda}^{\nu} \nabla_{\rho}(\sqrt{-g} f_R g^{\mu\rho})+\frac{1}{2} \delta_{\lambda}^{\mu} \nabla_{\rho}(\sqrt{-g} f_R g^{\nu\rho}) \Big] \delta \Gamma_{\nu\mu}^{\lambda} \nonumber  \ ,
\ena
where a total derivative (boundary) term has been discarded. This variation must satisfy $\delta_{\Gamma} S =0$. Noting that the contraction of the indices $\lambda$ and $\nu$ leads to $(1-n)\nabla_{\rho}(\sqrt{-g} f_R g^{\mu\rho})=0$, we conclude that the variation of the connection leads to
\be
\nabla_{\lambda} \left(\sqrt{-g} f_R g^{\mu\nu} \right) =0, \label{eq:connection}
\en

On the other hand, since the energy-momentum tensor of the matter is obtained as $T_{\mu\nu}=-\frac{2}{\sqrt{-g}}\frac{\delta (\sqrt{-g}L_m)}{\delta g^{\mu\nu}}$ (note that the matter Lagrangian, $L_m$, depends on the metric only) the field equations for $g_{\mu\nu}$ can be immediately obtained from the corresponding piece in (\ref{eq:var-total}) as

\be
f_R R_{\mu\nu}-\frac{f}{2}g_{\mu\nu}=\kappa^2 T_{\mu\nu}. \label{eq:metric}
\en

It is important to note that the assumption of independence between metric and connection has been crucial to obtain Eqs.(\ref{eq:metric}) and (\ref{eq:connection}). Had we followed the traditional approach of imposing  that $\Gamma_{\mu\nu}^{\lambda}$ is \emph{a priori} the Levi-Civita connection of $g_{\mu\nu}$, then we should have replaced $\delta \Gamma_{\nu\mu}^{\lambda}$ in (\ref{eq:con-var}) by
\begin{equation}
\delta \Gamma_{\nu\mu}^{\lambda}=\frac{g^{\lambda\rho}}{2}\left[\nabla_\mu \delta g_{\rho\nu}+\nabla_\nu \delta g_{\rho\mu}-\nabla_\rho \delta g_{\mu\nu}\right]  \ .
\end{equation}
In the usual metric approach, these terms must be integrated by parts and added to (\ref{eq:var-total}) in order to get the complete variation of the metric. As a result, one would have obtained a modification of (\ref{eq:metric}) of the form
\be
f_R R_{\mu\nu}-\frac{f}{2}g_{\mu\nu}+\nabla_\mu\nabla_\nu f_R-g_{\mu\nu}\Box f_R=\kappa^2 T_{\mu\nu}  \ , \label{eq:metric-Riemann}
\en
which contains higher-order derivatives of the metric through the terms $\nabla_\mu\nabla_\nu f_R$ and $\Box f_R$ stemming from the integration by parts of $\delta \Gamma_{\nu\mu}^{\lambda}\sim g^{\lambda\rho}\nabla_\mu \delta g_{\rho\nu}$.
The independence between metric and connection is thus the key to avoid these higher-order derivative terms in the field equations (\ref{eq:metric}) and (\ref{eq:connection}). For a comparison of the dynamics of $f(R)$ theories in both their  metric and Palatini formulations see \cite{Olmo05}.

To solve the system of equations (\ref{eq:metric}) and (\ref{eq:connection}) we take advantage of the fact that the connection equation (\ref{eq:connection}) can be formally written as

\be \label{eq:newconnection}
\nabla_{\lambda} \left(\sqrt{-h} h^{\mu\nu}\right)=0,
\en
where $h_{\mu\nu}$ is a symmetric rank-two tensor. Comparison between (\ref{eq:connection}) and (\ref{eq:newconnection}) leads to the relation

\be \label{eq:h}
h_{\mu\nu}=f_R^{\frac{2}{n-2}} g_{\mu\nu} \hspace{0.1cm}; \hspace{0.1cm} h^{\mu\nu}=f_R^{\frac{2}{2-n}} g^{\mu\nu},
\en
between $h_{\mu\nu}$ and $g_{\mu\nu}$, which means that they are related by a conformal transformation. On the other hand, contracting in Eq.(\ref{eq:metric}) with the metric $g^{\mu\nu}$ it follows that

\be \label{eq:trace}
Rf_R-\frac{n}{2}f=\kappa^2 T,
\en
where $T\equiv {T_\mu}^{\mu}$ is the trace of the energy-momentum tensor. This is an algebraic equation whose formal solution $R=R(T)$ generalizes the GR relation $R=-\kappa^2 T$ to the case of nonlinear $f(R)$ Lagrangian. The conformal factor relating $g_{\mu\nu}$ and $h_{\mu\nu}$ is thus a function of the matter, $f_R=f_R(R[T])$, which becomes a constant when $T=0$ (vacuum or traceless sources). Through a constant rescaling of the metric, one can then have  $g_{\mu\nu}=h_{\mu\nu}$ when $T=0$. It is also easy to see  that in vacuum the field equations  (\ref{eq:metric}) recover the GR dynamics with an effective cosmological constant, $G_{\mu\nu}=-\Lambda_{eff} g_{\mu\nu}$, where $\Lambda_{eff}\equiv (n-2)(f/4f_R)|_{T=0}$ depends on the specific form of the Lagrangian density chosen. This is a manifestation of the observed universality of Einstein's field equations in vacuum for Palatini theories \cite{Mauro} and implies that these theories do not propagate extra degrees of freedom. Therefore, in order to obtain modified dynamics, matter sources with $T \neq 0$ must be considered regardless of the number of space-time dimensions.

Having this in mind we now contract (\ref{eq:metric}) with $h^{\mu\nu}$, and arrange terms to obtain
\be \label{eq:Rmunu}
{R_\mu}^{\nu}(h)=\frac{1}{f_R^{\frac{2}{n-2}}} \left(\frac{f}{2} \delta_{\mu}^{\nu} +\kappa^2 {T_\mu}^{\nu} \right)
\en
which is a set of second-order Einstein-like field equations for $h_{\mu\nu}$. Note that since $R=R(T)$, the $f$ and $f_R$ terms on the right-hand side of (\ref{eq:Rmunu}) are functions of the matter. As $h_{\mu\nu}$ is algebraically related to $g_{\mu\nu}$ via the matter sources, the field equations for the latter are second-order as well, which is in agreement with our initial claim regarding the absence of higher-order field equations in Palatini $f(R)$ gravity in any dimension. Once a given matter-energy source is specified, the field equations (\ref{eq:Rmunu}), together with Eqs.(\ref{eq:h}) and (\ref{eq:trace}), provide a complete solution.

A comment on the degree of differentiability of the matter fields is now in order. In fact, since the conformal factor that relates the physical and auxiliary geometries depends on the matter fields, in order to have a smooth $g_{\mu\nu}$ geometry the function $f_R(R[T])$ must have, at least, smooth derivatives up to second order in the space-time coordinates. It is thus reasonable to wonder if higher-order derivatives of the matter fields might appear in these theories. Through a Hamiltonian analysis, it was found in \cite{Olmo:2011fh} that, in general, the field equations do involve higher-order {\it spatial} derivatives of the matter fields, though time derivatives remain second-order at most. Therefore, the Cauchy problem in these theories is as well-formulated as in GR. The well-posedness depends on the particular matter fields considered and it has been shown to be well-posed for a number of reasonable sources \cite{Capo}.

\subsection{Electrovacuum fields}

As we have discussed, in order to excite the dynamics of Palatini $f(R)$ gravity, one needs a matter-energy source with non-vanishing trace. A simple scenario is to consider the case of electromagnetic fields. In this sense, note that in four dimensions the Maxwell stress-energy tensor satisfies $T=0$, which forces the consideration of  non-linear theories of electrodynamics \cite{fR} in order to achieve modifications as compared to GR. For higher dimensions, however,  the Maxwell field satisfies $T \neq 0$ and does provide modified dynamics on its own. The action of Maxwell electrodynamics is given by
\be
S_m=-\frac{1}{16\pi} \int d^n x \sqrt{-g}F_{\mu\nu}F^{\mu\nu},
\en
where $F_{\mu\nu}=\partial_{\mu}A_{\nu}-\partial_{\nu}A_{\mu}$ is the field strength tensor of the vector potential $A_{\mu}$. Let us now assume a static, spherically symmetric line element

\be
ds^2=g_{tt}dt^2 -g_{rr} dr^2 -r^2 d\Omega_{n-2}^2,
\en
where $d\Omega_{n-2}^2=d\theta_1^2+\sum_{i=2}^{n-1}\prod_{j=1}^{i-2} \sin^2 \theta_{j} d\theta_{i}^2$ is the metric on the unit ($n-2$) sphere. In this line element, from the Maxwell field equations, $\nabla_{\mu}F^{\mu\nu}=0$, one finds that the  unique non-vanishing component of a spherically symmetric, electrically charged field  reads $F^{tr}=\frac{q}{r^{(n-2)} \sqrt{-g_{tt}g_{rr}}}$, where $q$ is an integration constant identified as the electric charge. Remarkably, the invariant

\be \label{eq:X}
X=-\frac{1}{2} F_{\mu\nu}F^{\mu\nu}=\frac{q^2}{r^{2(n-2)}},
\en
does not depend explicitly on the $g_{tt}$ and $g_{rr}$ components of the metric, which simplifies the calculations. This allows us to write the energy-momentum tensor of the electromagnetic field for these solutions as

\bea \label{eq:Tmunu}
{T_\mu}^{\nu}&=&-\frac{1}{4\pi} \left( {F_\mu}^{\alpha} {F_\alpha}^{\nu}-\frac{1}{4} \delta_{\mu}^{\nu}F_{\alpha\beta}F^{\alpha \beta} \right)  \nonumber \\
&=&\frac{X}{4\pi}
\left(
\begin{array}{cc}
-\hat{I}_{2\times 2}&  \hat{0}_{(n-2) \times 2} \\
\hat{0}_{2 \times (n-2)} & \hat{I}_{(n-2) \times (n-2)} \\
\end{array}
\right), \label{eq:em}
\ena
where $\hat{I}_{a \times b}$ and $\hat{0}_{a \times b}$ are the $a \times b$ dimensional identity and zero matrices, respectively. From (\ref{eq:em}) we explicitly read the non-vanishing trace $T= \frac{(n-4)q^2}{4\pi r^{2(n-2)}}$ for $n \neq 4$.

\subsection{Computation of the five-dimensional geometrical objects}

To solve the field equations (\ref{eq:Rmunu}) we must compute the objects appearing on the left-hand side. As later we shall focus on the properties of five-dimensional solutions, let us propose the following ansatz for the metric $h_{\mu\nu}$
\bea \label{eq:ansatz}
d\tilde{s}^2&=&-A(x)e^{2\xi(x)}dt^2+\frac{1}{A(x)} dx^2 \nonumber \\
&+&\tilde{r}^2(x)(d\theta^2 +\sin^2 \theta d\phi^2 + \sin^2 \theta \sin^2 \phi  d\alpha^2),
\ena
where $\tilde{r}^2(x)$ is, in general, a function of the coordinate $x$. For the problem at hand, however, the choice $\tilde{r}(x)=x$ turns out to be a consistent one. Assuming this simplification, the computation of the metric components ${R_\mu}^{\nu}(x)$ (we use the xAct package of Mathematica \cite{JM}) leads to
\bea
{R_t}^t&=&-\frac{1}{2x} [3A_x (1+x\xi_x)+xA_{xx} \nonumber \\
&+&2A(3\xi_x + x(\xi_x^2 + \xi_{xx}))] \\
{R_x}^x&=&-\frac{1}{2x} [3A_x (1+x\xi_x)+xA_{xx} \nonumber \\
&+&2A x(\xi_x^2 + \xi_{xx}))] \\
{R_\theta}^{\theta}&=&{R_\theta}^{\theta}={R_\alpha}^{\alpha}=\frac{1}{x^2}[2(1-A)-xA_x-A x \xi_x] \label{eq:Rtheta}
\ena
Using the symmetry of the matter-energy source, ${T_t}^t={T_x}^x$, from (\ref{eq:Rmunu}) it follows that $\xi_x=0$. Therefore, we may set $\xi=0$ by a redefinition of the time coordinate without loss of generality. On the other hand, the component (\ref{eq:Rtheta}) is appropriately written using a suitable mass ansatz in five dimensions, $A=1-2M(x)/x^2$, in terms of which (\ref{eq:Rtheta}) reads

\be \label{eq:Rthetatheta}
{R_\theta}^{\theta}(h)=2M_x/x^3,
\en
which provides a solution once the corresponding component of the right-hand-side of (\ref{eq:Rmunu}) is given, for which the gravity Lagrangian $f(R)$ must be specified. In the following section we shall consider two examples.

\section{Some models}

For the sake of simplicity, in this work we shall restrict ourselves to the study of polynomial models of the form

\be \label{eq:fR}
f(R)=R+\alpha R^d,
\en
where $d$ is a constant and $\alpha$ a parameter. From the trace equation (\ref{eq:trace}) we obtain the relation $R=R(T)$ as

\be \label{eq:trace1}
R\left(\frac{2-n}{2} \right)+\alpha R^d \left(\frac{2d-n}{2} \right)=\kappa^2 T,
\en
whose explicit resolution must be done case-by-case.

\subsection{$f(R)=R+\alpha R^2$}

A natural choice is that of a quadratic Lagrangian, namely, $d=2$. For dimensional consistency, $\alpha$ has dimensions of length squared. In this case the equation for the trace (\ref{eq:trace}) is solved as
\be \label{eq:Rf1}
\alpha R=-\frac{(n-2)}{2 (n-4)} \left[ 1-\sqrt{1-\frac{8(n-4)\kappa^2 \alpha T}{(n-2)^2}} \right] \ ,
\en
which indicates that $R$ is negative if $\alpha>0$ and positive if $\alpha<0$. Note that  the term under the square root might become negative if $r$ becomes smaller than $r_\alpha^{2(n-2)}=\frac{8\alpha (n-4)\kappa^2 q^2}{(n-2)^2}$. However, one can see that $r_\alpha$ is smaller than the point $r_c$ where $f_R$ vanishes, which occurs at $r_c^{2(n-2)}=\frac{8\alpha \kappa^2 q^2}{n}$, where $\alpha R=-1/2$. As in other cases already studied in four dimensions \cite{or12}, the point where $f_R=0$ sets the location of a minimum in the function $r^2(x)$, which prevents the square root in (\ref{eq:Rf1}) from becoming complex. Such a minimum signals the existence of a wormhole throat. Determining the behaviour of the metric functions at the minimum $r_c$ of the radial coordinate is, consequently, a key aspect in the characterization of solutions in these theories. In order to find the explicit relation between the function $r^2(x)$ and the coordinate $x$, let us write the line element for the metric $g_{\mu\nu}$ as
\be \label{eq:ansatzg}
ds^2=-B(x)dt^2+C(x)dx^2+r^2(x) d\Omega^2_{n-3}.
\en
It is important to remember that the line elements for  $g_{\mu\nu}$ in (\ref{eq:ansatzg}) and $h_{\mu\nu}$ in (\ref{eq:ansatz}) are conformally related according to (\ref{eq:h}). This means that $r^2(x)=f_R^{\frac{2}{2-n}} x^2$. Therefore, computing $f_R$ and using Eq.(\ref{eq:Rf1}) we find
\be
x^2=r^2 \left[\frac{(n-2)}{(n-4)}\sqrt{1-\left(\frac{r_{\alpha}}{r} \right)^{2(n-2)}} - \frac{2}{(n-4)} \right]^{\frac{2}{n-2}} \ .
\en
Though the resolution of the field equations for this problem is certainly possible, the nonlinearity of the equation (\ref{eq:Rf1}) that determines the new objects appearing on the right-hand-side of the field equations (\ref{eq:Rmunu}) prevents the obtention of a clear analytical solution. This example was taken here to stress the relevance of the relation between the two-spheres of the geometries $g_{\mu\nu}$. However, as the aim of this paper is to illustrate the Palatini method in higher-dimensions we shall now choose an analytically tractable model.

\subsection{$f(R)=R+\beta |R|^{n/2}$}

The reason to consider this model (where $\beta$ is a parameter) is its technical simplicity: the trace equation is solved as
\be \label{eq:RT}
R=\frac{2}{(2-n)} \kappa^2 T \ ,
\en
which coincides with the GR expression and does not depend on the $\beta$ parameter.
The linearity of the Ricci scalar in the trace provides a simple expression of the $f(R)$ Lagrangian in terms of the matter sources as
\be
f(R)=\frac{2\kappa^2 T}{(2-n)} \left[1+\beta \left|\frac{2\kappa^2 T}{2-n} \right|^{\frac{n-2}{2}} \right].
\en
To obtain analytical solutions we shall focus on the $n=5$ dimensional case, for which we have $T=q^2/(4\pi r^6)$ and $R=-\frac{2}{3} \frac{r_q^4}{r^6}$, where we have defined the charge scale $r_q^4\equiv\kappa^2 q^2/(4\pi)$. These expressions and that of the energy-momentum tensor of electromagnetic field in (\ref{eq:Tmunu}) allow us to compute the right-hand side of the field equations (\ref{eq:Rmunu}). Therefore, using (\ref{eq:Rthetatheta}) we obtain
\be
{R_\theta}^{\theta}=\frac{2M_x}{x^3}=\frac{r_q^4}{f_R^{5/3} r^6} \left[1+\frac{1}{3} \left(1+\beta \left(\frac{2}{3} \right)^{5/2} \frac{r_q^6}{r^9}\right) \right].
\en
To integrate this equation we need the relation $x=rf_R^{1/3}$ between the spherical sectors of the line elements of $h_{\mu\nu}$ and $g_{\mu\nu}$, as follows from (\ref{eq:h}). It is easy to see that this implies $dx/dr=f_R^{1/3} \left[1+\frac{r}{3} \frac{f_{RR}}{f_R} R_r \right]$, where the explicit expression for $R_r$ follows from (\ref{eq:RT}). With this we can finally write
\be \label{eq:Mr1}
M_r=\frac{r^3}{2f_R^{1/3}} \left(1+\frac{r}{3} \frac{f_{RR}}{f_R} R_r \right) \left(\frac{r_q^4}{r^6}+\frac{ f(R)}{2} \right).
\en
The explicit expressions for $f(r)$ and $f_R(r)$ follow immediately as

\be
f(r)=-\frac{2}{3} \frac{r_q^4}{r^6} \left[1-\frac{2\tilde{\beta}}{5}\frac{r_q^6}{r^9} \right] \ ; \ f_R=1-\tilde{\beta} \frac{r_q^6}{r^9},
\en
where we have defined a new parameter $\tilde{\beta}=5\beta(2/3)^{3/2}/2$. Note that, on dimensional grounds, $\tilde{\beta}$ represents a cubic length.  If we interpret the non-linear curvature corrections as having a quantum-gravitational origin, then $\tilde{\beta}\sim l_P^3$, where $l_P$ is the Planck length. Using the fact that $f_{RR}R_r=9\tilde{\beta} \frac{r_q^6}{r^{10}}$ and defining a new scale $r_c^9 \equiv \tilde{\beta} r_q^6$, we can finally write Eq.(\ref{eq:Mr1}) for our theory as
\be
M_r=\frac{r_q^4}{3r^3} \frac{\left[1+2\left(\frac{r_c}{r} \right)^9 \right] \left[1+\frac{1}{5}\left(\frac{r_c}{r}\right)^9\right]}{\left[1-\left(\frac{r_c}{r} \right)^9 \right]^{4/3}}.
\en
Since for zero charge we have $M=M_0=$constant,  this expression can be more conveniently written in terms of a function $G(z)$ defined as
\be
\frac{M(z)}{M_0}=1+\delta_1 G(z) \ ,
\en
where we have introduced the new variable $z=r/r_c$ and the constant $\delta_1 \equiv \frac{r_q^4}{3M_0 r_c^2}$. The function $G(z)$ satisfies
\be \label{eq:Gz}
G_z=\frac{1}{z^3} \frac{\left[1+\frac{2}{z^9} \right] \left[1+\frac{1}{5z^9} \right]}{\left(1-\frac{1}{z^9} \right)^{4/3}},
\en
and contains the electromagnetic contribution. This function admits an immediate integration as

\be \label{eq:G(z)}
G(z)= \epsilon + \frac{1-25 z^9-30 z^9 (1-z^9)^{1/3} \, _2F_1\left(\frac{1}{9},\frac{1}{3};\frac{10}{9};z^9\right)}{20 z^8 (z^9-1)^{1/3}},
\en
where $_2F_1$ is a hypergeometric function and $\epsilon = 3 (-1)^{2/9} \Gamma \left(\frac{2}{9}\right) \Gamma \left(\frac{10}{9}\right)/(2 \Gamma \left(\frac{1}{3}\right))$ is a constant needed to recover the right five-dimensional GR behaviour at $z \gg 1$. Indeed, a series expansion in this region gives

\be
G(z) \approx -\frac{1}{2 z^2}-\frac{53}{165z^{11}}+O\left(\frac{1}{\text{z}}\right)^{13},
\en
where the leading-order term displays the expected GR behaviour and the corrections are largely suppressed as $1/z^9= \tilde{\beta} r_q^6/r^9\ll 1$ for any $r>l_P\sim \tilde{\beta}^{1/3}$.

\begin{figure}[h]
\includegraphics[width=0.45\textwidth]{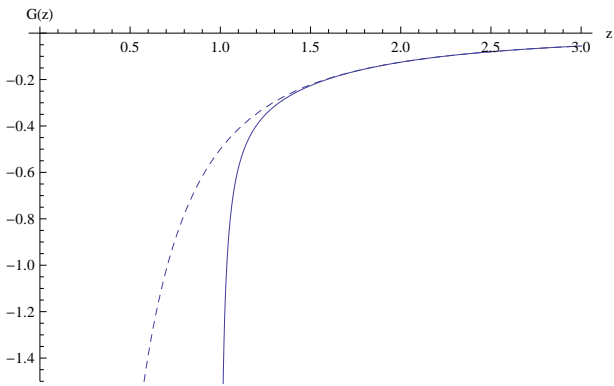}
\caption{Behaviour of the function $G(z)$ (solid) in Eq.(\ref{eq:G(z)}), as compared to the GR case (dashed). Far from the center, $z \gg 1$, the solution quickly converges to that of GR, $G(z)=-1/(2z^2)$, but in the other region of interest it undergoes relevant modifications as a core of non-vanishing radius $z=1$ arises. At this core curvature invariants diverge (see Eq.(\ref{eq:Kret}) below).
\label{fig:1}}
\end{figure}

We expect the effects of the correcting term in the Lagrangian to produce deviations from the five-dimensional GR solutions around the region $z=1$ (see Figure \ref{fig:1}). Expanding (\ref{eq:G(z)}) in this region we obtain

\bea \label{eq:G(z)small}
G(z) &\approx& -\frac{2 \sqrt[3]{3}}{5 \sqrt[3]{z-1}} \nonumber \\
&+&\frac{3 (-1)^{2/9} \Gamma \left(\frac{2}{9}\right) \Gamma \left(\frac{7}{9}\right)-2 \sqrt[3]{-1} \sqrt{3} \pi  \Gamma \left(\frac{10}{9}\right)}{2 \Gamma \left(\frac{1}{3}\right) \Gamma \left(\frac{7}{9}\right)} \nonumber \\
&-& \left(\frac{1}{20\ 3^{2/3}}-\frac{\pi }{\sqrt[6]{3} \Gamma \left(\frac{1}{3}\right) \Gamma \left(\frac{5}{3}\right)}\right) (z-1)^{2/3} \nonumber \\
&+&O\left(z-1 \right)^1,
\ena

\begin{figure}[h]
\includegraphics[width=0.45\textwidth]{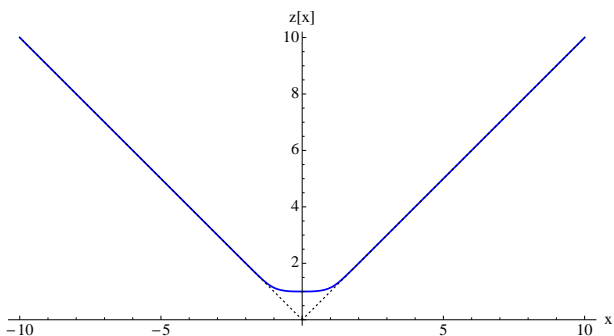}
\caption{Representation of the radial function $z=z(x)$. Note the rapid transition from linearity as $x\to 0$ is approached. The surface $x=0$ represents a minimum of the area function $r^2(x)=r_c^2 z^2(x)$ and can be interpreted as the throat of a wormhole.
\label{fig:2}}
\end{figure}

\noindent where $\Gamma[a]$ is Euler's gamma function. This expansion reveals that $G(z)$ is (slowly) divergent around $z=1$. To determine the impact of this divergence on the geometry, we note that the physical metric component in the line element (\ref{eq:ansatzg}) is completely determined via the conformal transformation (\ref{eq:h}) as
\be
B(z)=\frac{1}{f_R^{2/3}} \left(1-\frac{2M(z)}{r_c^2 z^2 f_R^{2/3}} \right).
\en
Expanding in series around $z=1$ this function behaves as
\bea
B & \approx& \frac{4 M_0 \delta_1}{45 (3)^{1/3} r_c^2 (z-1)^{5/3}} - \frac{2M_0(\delta_1 + C_1)}{9 3^{2/3} r_c^2 (z-1)^{4/3}} \nonumber \\
&+& \frac{90\sqrt{3}r_c^2 + C_2 M_0 \delta_2}{270 3^{5/6}r_c^2 (z-1)^{2/3}} + O(z-1)^{1/3},
\ena
where $C_1$ and $C_2$ are two constants whose explicit form is too large and of no particular interest for our purposes. The leading term in this expansion diverges as $1/(z-1)^{5/3}$. This allows to compute the behaviour of the curvature invariants at $z=1$, in particular, the Kretchsmann behaves there as

\be \label{eq:Kret}
R_{\alpha\beta\gamma\delta}R^{\alpha\beta\gamma\delta} \simeq \frac{1024 \delta_1^2 M_0^2}{65613^{2/3} r_c^8 (z-1)^{22/3}}+ O\left(\frac{1}{(z-1)^{20/3}} \right),
\en
which is thus divergent, regardless of the specific value of the constant $\delta_1$.

We would like to point out that the presence of a curvature divergence at $x=0$, where the minimum value $z=1$ is reached (see Fig.\ref{fig:2}), is not an obstacle for the existence of a wormhole (in fact, the geometry does not dictate the topology of a space). Our theory is defined by a gravity action coupled to a {\it sourceless} electric field and the resulting dynamics forces the radial function $r^2(x)$ to have a minimum at $x=0$ ($r=r_c$ or $z=1$). The non-zero electric flux through the $x=0$ surface allows to define the electric charge $q$ that characterizes our solutions in purely topological terms \cite{Misner-Wheeler}. Therefore, rather than being generated by a point-like source, the charge of our solutions is a topological property, which allows to interpret these solutions as geons in Wheeler's sense \cite{Geons}. We note that similar solutions but with completely smooth curvature scalars have been found in four-dimensional Palatini theories \cite{or12,ors,Guendelman:2013sca}. Moreover, it has been recently shown that  wormhole solutions of this kind can be dynamically generated out of Minkowski space by means of an ingoing stream of charged null particles \cite{Dynamical}.

Though different aspects such as the classification of the solutions in terms of their horizons, or their thermodynamical aspects could be certainly studied, we shall not keep going with the geometric analysis of these solutions as this model was introduced as an example that illustrates the analytical tractability of the second-order equations of Palatini $f(R)$ theories in five dimensions.  The important point of bringing this example is to show that the Palatini approach provides a new framework to consistently study extensions of GR with new curvature couplings and second-order equations in arbitrary dimension. Perturbations of these solutions will be considered elsewhere.

\section{Summary and outlook}

It is widely known that in the standard metric (or Riemannian) approach the addition in the action of higher-curvature terms with arbitrary coefficients in $d\ge 4$ dimensions breaks the second-order character of the field equations. The metric-affine or Palatini formulation, however, naturally avoids this shortcoming in a large family of models. In this paper we have illustrated this fact with the particular case of $f(R)$ theories, though this procedure can be easily extended to theories containing powers of the Ricci tensor (and possibly the Kretchsmann as well), which would extend the scope of these methods.

The second-order character of the Palatini field equations makes it possible to obtain exact solutions regardless of the dimension. We have explicitly run this idea by using the fact that, though the independent connection is not metric-compatible, it is compatible with a new metric, $h_{\mu\nu}$, algebraically related to the metric $g_{\mu\nu}$ via the (trace of the) energy-momentum tensor of the matter. The field equations in terms of $h_{\mu\nu}$ can be cast in Einstein-like form, which simplifies their analysis and resolution. This procedure provides a full solution for a given matter-energy source. It is worth mentioning that in theories beyond $f(R)$ the algebraic relation between $h_{\mu\nu}$ and $g_{\mu\nu}$ transcends the conformal case but its determination is, in principle, possible in many cases of interest \cite{Olmo:2011uz,Olmo:2009xy}.

We have studied a scenario in which an electromagnetic (Maxwell) field in $n>4$ dimensions sources a family of polynomial $f(R)$ theories. Unlike in the $n=4$ case, where the Maxwell stress-energy tensor is traceless, in $n=5$ the trace of electromagnetic field is non-vanishing,  which allows to probe the modified gravitational dynamics. We have successfully obtained exact solutions to the field equations of the model $f(R)=R+\beta |R|^{5/2}$, which was chosen by its analytical tractability, and studied how the region close to $r=0$ is modified by the new high-energy dynamics. The main novelty is the fact that these solutions do not extend all the way down to $r=0$ but, instead, a sphere of minimum area arises as a consequence of the new dynamics and the conformal relation between the $2$-spheres of the $h_{\mu\nu}$ and $g_{\mu\nu}$ geometries. The existence of this {\it core}, which is a manifestation of the existence of a wormhole, seems to be a generic prediction of Palatini theories sourced by electrovacuum fields, as it has been found to arise in the context of four-dimensional $f(R)$ coupled to Born-Infeld electrodynamics \cite{fR}, and also in quadratic gravity \cite{or12} and Born-Infeld-like gravity \cite{ors}. In the present case, curvature invariants for the particular model considered diverge at the core in all cases, a situation similar  to that found in the four-dimensional $f(R)$ context \cite{fR} but which can be cured in other extensions beyond the $f(R)$ scenario \cite{or12,ors,Guendelman:2013sca}.

The important lesson that follows from our discussion is that a foundational aspect of gravity, namely, whether the underlying structure of space-time is Riemannian or not, has a great influence on both the mathematical and physical aspects of the corresponding theory. The Palatini approach to $f(R)$ gravities shows that one can add to the action as many new couplings in the gravitational field as desired without spoiling the second-order character of the field equations. One may wonder whether the addition of other curvature invariants such as $R_{\mu\nu}R^{\mu\nu}$ or $R_{\alpha\beta\gamma\delta}R^{\alpha\beta\gamma\delta}$ keeps the second-order character. We have already checked that this is so in the case of four-dimensional Born-Infeld  \cite{ors,Deser} and quadratic gravity \cite{Guendelman:2013sca,or12}, which contain non-trivial corrections on the Ricci-squared invariant.

We stress that the vacuum field equations of Palatini $f(R)$ (and further extensions) boil down to those of GR plus a cosmological constant. Alternatively a cosmological constant term can be directly added to the action in a standard way, or generated through non-linear corrections in the electromagnetic field \cite{Guendelman:2013sca}, neither of these ways spoiling the second-order character of the field equations. In summary, the Palatini approach provides an interesting framework to explore new domains of the AdS/CFT correspondence and to potentially broaden the class of CFTs that can be studied using holographic methods. A better understanding of the theory of quantized fields in these backgrounds is, however, necessary to  fully understand how this correspondence manifests itself in such scenarios. These are aspects to be explored elsewhere.

\section*{Acknowledgments}

D.B. and L.L. would like to thank CAPES and CNPq for financial support. G.J.O. is supported by the Spanish grant FIS2011-29813-C02-02, the Consolider Program CPANPHY-1205388, the JAE-doc program of the Spanish Research Council (CSIC), and the i-LINK0780 grant of CSIC. D.R.-G. is supported by CNPq through project No. 561069/2010-7. The authors also acknowledge funding support of CNPq project No. 301137/2014-5.



\end{document}